\documentclass[a4paper,12pt]{article}
\usepackage{amssymb,amsmath,amsfonts}
\usepackage{fullpage}
\usepackage{graphicx}
\usepackage[latin1]{inputenc}

\begin{document}

\title{Explicit expressions for moments of the beta Weibull distribution}
\date{}
\author{Gauss M. Cordeiro$^{a,}$\footnote{Corresponding author. E-mail: gausscordeiro@uol.com.br},~ Alexandre B. Simas$^{b,}$\footnote{E-mail: alesimas@impa.br}~ and Borko D. Sto\v si\' c$^{a,}$\footnote{E-mail: borko@ufpe.br}\\\\
\centerline{\small{
$^a$Departamento de Estat\' \i stica e Inform\' atica, 
Universidade Federal Rural de Pernambuco,
}}\\
\centerline{\small{
Rua Dom Manoel de Medeiros s/n, Dois Irm\~ aos,
52171-900 Recife-PE, Brasil}}\\
\centerline{\small{
$^b$Associa\c{c}\~ao Instituto Nacional de Matem\'atica Pura e Aplicada, IMPA,}}\\
\centerline{\small{
Estrada D. Castorina, 110, Jd. Bot\^anico, 22460-320, Rio de Janeiro-RJ, Brasil}}}
\maketitle
\begin{abstract}
The beta Weibull distribution was introduced by Famoye et al. (2005)
and studied by these authors. However, they do not give explicit
expressions for the moments. We now derive explicit closed form
expressions for the cumulative distribution function and for the
moments of this distribution. We also give an asymptotic expansion
for the moment generating function. Further, we discuss maximum
likelihood estimation and provide formulae for the elements of the
Fisher information matrix. We also de\-mons\-trate the usefulness of
this distribution on a real data set.
\end{abstract}

\section{Introduction}

The Weibull distribution is a popular distribution widely used for
analyzing lifetime data. We work with the beta Weibull (BW)
distribution because of the wide applicability of the Weibull
distribution and the fact that it extends some recent developed
distributions. This generalization may attract wider application in
reliability and biology. We derive explicit closed form expressions
for the distribution function and for the moments of the BW
distribution. An application is illustrated to a real data set with
the hope that it will attract more applications in reliability and
biology, as well as in other areas of research.

The BW distribution stems from the following general class: if $G$
denotes the cumulative distribution function (cdf) of a random
variable then a generalized class of distributions can be defined by
\begin{equation}\label{betadist}
F(x) = I_{G(x)}(a,b)
\end{equation}
for $a>0$ and $b>0$, where
$$I_y(a,b) = \frac{B_y(a,b)}{B(a,b)}=\frac{\int_0^y w^{a-1}(1-w)^{b-1}dw}{B(a,b)}$$
is the incomplete beta function ratio, $B_y(a,b)$ is the incomplete
beta function, $B(a,b)=\Gamma(a)\Gamma(b)/\Gamma(a+b)$ is the beta
function and $\Gamma(.)$ is the gamma function. This class of
generalized distributions has been receiving increased attention
over the last years, in particular after the recent works of Eugene
et al. (2002) and Jones (2004). Eugene et al. (2002) introduced what
is known as the beta normal distribution by taking $G(x)$ in
(\ref{betadist}) to be the cdf of the normal distribution with
parameters $\mu$ and $\sigma$. The only properties of the beta
normal distribution known are some first moments derived by Eugene
et al. (2002) and some more general moment expressions derived by
Gupta and Nadarajah (2004). More recently, Nadarajah and Kotz (2004)
were able to provide closed form expressions for the moments, the
asymptotic distribution of the extreme order statistics and the
estimation procedure for the beta Gumbel distribution. Another
distribution that happens to belong to (\ref{betadist}) is the
log$\,F$ (or beta logistic) distribution, which has been around for
over 20 years (Brown et al., 2002), even if it did not originate
directly from (\ref{betadist}).

While the transformation (\ref{betadist}) is not analytically
tractable in the general case, the formulae related with the BW turn
out manageable (as it is shown in the rest of this paper), and with
the use of modern computer resources with analytic and numerical
capabilities, may turn into adequate tools comprising the arsenal of
applied statisticians. The current work represents an advance in the
direction traced by Nadarajah and Kotz (2006), contrary to their
belief that some mathematical properties of the BW distribution are
not tractable.

Thus, following (\ref{betadist}) and replacing $G(x)$ by the cdf of a Weibull
distribution with parameters $c$ and $\lambda$, we obtain the cdf of the BW distribution
\begin{equation}\label{bwdist}
F(x) = I_{1 - {\rm exp}\{-(\lambda x)^c\}}(a,b)
\end{equation}
for $x>0$, $a>0$, $b>0$, $c>0$ and $\lambda>0$. The corresponding probability density function (pdf) and the
hazard rate function associated with (\ref{bwdist}) are:
\begin{equation}\label{bwpdf}
f(x) = \frac{c\lambda^c}{B(a,b)}x^{c-1} {\rm exp}\{-b(\lambda x)^c\}[1-{\rm exp}\{-(\lambda x)^c\}]^{a-1} ,
\end{equation}
and
\begin{equation}\label{bwhaz}
\tau(x) = \frac{c\lambda^c x^{c-1} {\rm exp}\{-b(\lambda x)^c\}[1-{\rm exp}\{-(\lambda x)^c\}]^{a-1}}{ B_{1 - {\rm exp}\{-(\lambda x)^c\}}(a,b)} ,
\end{equation}
respectively. Simulation from (\ref{bwpdf}) is easy: if $B$ is a
random number following a beta distribution with parameters $a$ and
$b$ then $X = \{-{\rm log}(1-B)\}^{1/c}/\lambda$ will follow a BW
distribution with parameters $a, b, c$ and $\lambda$. Some
mathematical properties of the BW distribution are given by Famoye
et al. (2005) and Lee et al. (2006).

\begin{figure}[!h]\includegraphics[width=5.0in]{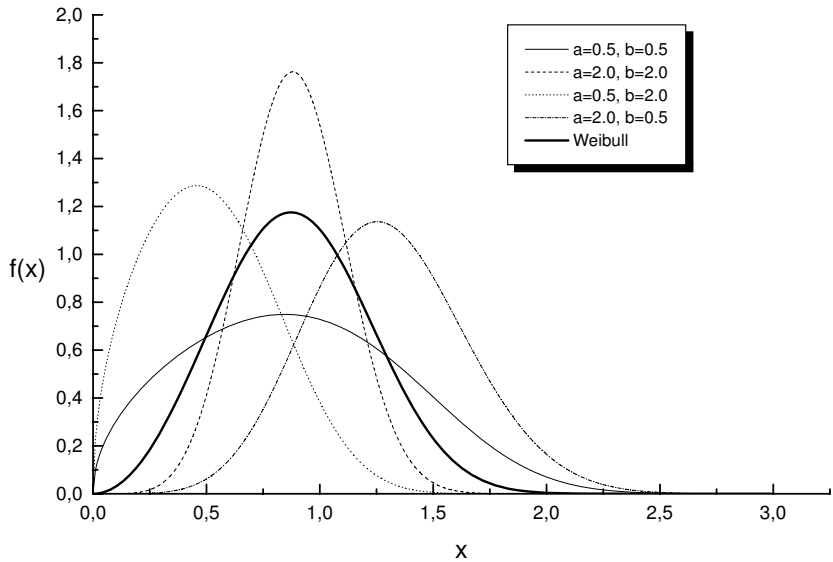}
\caption{\label{fig1}
The probability density function (\ref{bwpdf})
of the BW distribution, for several values
of parameters $a$ and $b$
}
\end{figure}

\begin{figure}[!h]\includegraphics[width=5.0in]{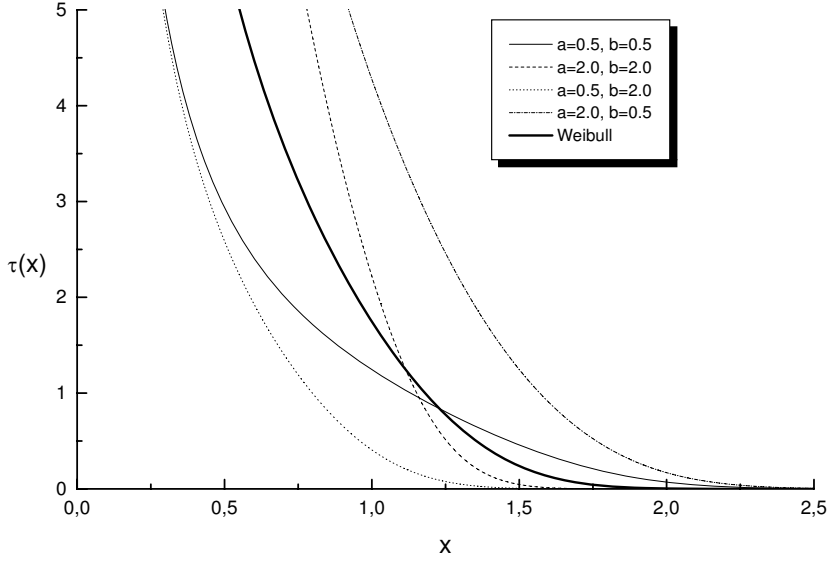}
\caption{\label{fig2}
The hazard function (\ref{bwhaz})
of the BW distribution, for several values
of parameters $a$ and $b$
}
\end{figure}

Graphical representation of equations (\ref{bwpdf}) and
(\ref{bwhaz}) for some choices of parameters $a$ and $b$, for fixed
$c=3$ and $\lambda=1$, are given in Figures \ref{fig1} and
\ref{fig2}, respectively. It should be noted that a single Weibull
distribution for the particular choice of the parameters $c$ and
$\lambda$ is here generalized by a family of curves with a variety
of shapes, shown in these figures.

The rest of the paper is organized as follows. In Section 2, we
obtain some expansions for the cdf of the BW distribution, and point
out some special cases that have been considered in the literature.
In Section 3, we derive explicit closed form expressions for the
moments and present skewness and kurtosis for different parameter
values. Section 4 gives an expansion for its moment generating
function. In Section 5, we discuss the maximum likelihood estimation
and provide the elements of the Fisher information matrix. In
Section 6, an application to real data is presented, and finally, in
Section 7, we provide some conclusions. In the appendix, two
identities needed in Section 3 are derived.

\section{Expansions for the distribution function}

The BW distribution is an extended model to analyze more complex
data and generalizes some recent developed distributions. In
particular, the BW distribution contains the exponentiated Weibull
distribution (for instance, see Mudholkar et al., 1995, Mudholkar
and Hutson, 1996, Nassar and Eissa, 2003, Nadarajah and Gupta, 2005
and Choudhury, 2005) as special cases when $b=1$. The Weibull
distribution (with parameters $c$ and $\lambda$) is clearly a
special case for $a=b=1$. When $a=1$, (\ref{bwpdf}) follows a
Weibull distribution with parameters $\lambda \, b^{1/c}$ and $c$.
The beta exponential distribution (Nadarajah and Kotz, 2006) is also
a special case for $c=1$.

In what follows, we provide two simple formulae for the cdf
(\ref{bwdist}), depending on whether the parameter $a>0$ is real
non-integer or integer, which may be used for further analytical or
numerical analysis. Starting from the explicit expression for the
cdf (\ref{bwdist})
$$F(x) = \frac{c\lambda^c}{B(a,b)} \int_0^x  y^{c-1} \exp\{-b(\lambda y)^c\}[1-\exp\{-(\lambda y)^c\}]^{a-1} dy,$$
the change of variables $(\lambda y)^c = u$ yields
$$F(x) = \frac{1}{B(a,b)} \int_0^{(\lambda x)^c} e^{-bu} (1-e^{-u})^{a-1} du.$$
If $a>0$ is real non-integer we have
\begin{equation}\label{expreal}
(1-z)^{a-1} = \sum_{j=0}^\infty \frac{(-1)^j \Gamma(a)
z^j}{\Gamma(a-j)j!}.
\end{equation}
It follows that
\begin{eqnarray*}
F(x)&=& \frac{1}{B(a,b)} \int_0^{(\lambda x)^c} e^{-bu} \sum_{j=0}^\infty \frac{(-1)^j \Gamma(a) e^{-ju}}{\Gamma(a-j)j!} du\\
&=& \frac{1}{B(a,b)} \sum_{j=0}^\infty \frac{(-1)^j \Gamma(a)}{\Gamma(a-j)j!} \int_0^{(\lambda x)^c} e^{-(b+j)u} du\\
&=& \frac{1}{B(a,b)} \sum_{j=0}^\infty \frac{(-1)^j \Gamma(a)}{\Gamma(a-j)j! (b+j)} \{ 1- e^{-(b+j)(\lambda x)^c}\}.
\end{eqnarray*}
Finally, we obtain
\begin{equation}\label{bwdistexp}
F(x) = \frac{\Gamma(a+b)}{\Gamma(b)}\sum_{j=0}^\infty \frac{(-1)^j\{ 1- e^{-(b+j)(\lambda x)^c}\}}{\Gamma(a-j)j! (b+j)}.
\end{equation}
For positive real non-integer $a$, the expansion (\ref{bwdistexp})
may be used for further analytical and/or numerical studies. For
integer $a$ we only need to change the formula used in
(\ref{expreal}) to the binomial expansion to give
\begin{equation}\label{bwdistexpint}
F(x) = \frac{1}{B(a,b)}\sum_{j=0}^{a-1} \binom{a-1}{j} \frac{(-1)^j\{ 1- e^{-(b+j)(\lambda x)^c}\}}{(b+j)}.
\end{equation}

When both $a$ and $b=n-a+1$ are integers,
the relation of the incomplete beta function to the binomial expansion gives

\begin{equation}\label{part1}
F(x) = \sum_{j=a}^n \binom{n}{j} [1-{\rm exp}\{-(\lambda x)^c\}]^{j}{\rm exp}\{-(n-j)(\lambda x)^c\}.
\end{equation}

It can be found in the Wolfram Functions Site\footnote{\tt http://functions.wolfram.com/GammaBetaErf/BetaRegularized/03/01/}
that for integer $a$
$$I_y(a,b) = 1-\frac{(1-y)^b}{\Gamma(b)} \sum_{j=0}^{a-1} \frac{\Gamma(b+j)y^j}{j!},$$
and for integer $b$,
$$I_y(a,b) = \frac{y^a}{\Gamma(a)}\sum_{j=0}^{b-1} \frac{\Gamma(a+j) (1-y)^j}{j!}.$$

Then, if $a$ is integer, we have another equivalent form for (\ref{bwdistexpint})
\begin{equation}\label{part2}
F(x) = 1 - \frac{{\rm exp}\{-b(\lambda x)^c\}}{\Gamma(b)}\sum_{j=0}^{a-1} \frac{\Gamma(b+j)}{j!} [1 - {\rm exp}\{-(\lambda x)^c\}]^{j}.
\end{equation}
For integer values of $b$, we have
\begin{equation}\label{part3}
F(x) = \frac{[1-{\rm exp}\{-(\lambda x)^c\}]^a}{\Gamma(a)} \sum_{j=0}^{b-1}\frac{\Gamma(a+j)}{j!}{\rm exp}\{-j(\lambda x)^c\}.
\end{equation}
Finally, if $a=1/2$ and $b=1/2$, we have
\begin{equation}\label{part4}
F(x) = \frac{2}{\pi} {\rm arctan}\sqrt{{\rm exp}\{(\lambda x)^c\} - 1}.
\end{equation}
The particular cases (\ref{part2}) and (\ref{part3}) were discussed
generally by Jones (2004), and the expansions
(\ref{bwdistexp})-(\ref{part4}) reduce to Nadarajah and Kotz's
(2006) results for the beta exponential distribution by setting
$c=1$. Clearly, the expansions for the BW density function are
obtained from (\ref{bwdistexp}) and (\ref{bwdistexpint}) by simple
differentiation. Hence, the BW density function can be expressed in
a mixture form of Weibull density functions.

\section{Moments}

Let $X$ be a BW random variable following the density function
(\ref{bwpdf}). We now derive explicit expressions for the moments of
$X$. We now introduce the following notation (for any real $d$ and
$a$ and $b$ positive)
\begin{equation}\label{sdba}
S_{d,b,a} = \int_0^\infty x^{d-1}{\rm exp}(-bx)\{1-{\rm
exp}(-x)\}^{a-1} dx.
\end{equation}
The change of variables $x = -{\rm log}(z)$ immediately yields
$S_{1,b,a} = B(a,b)$. On the other hand, the change of variables $x
= (\lambda y)^c$ gives the following relation
\begin{equation}\label{rel1}
\int_0^\infty y^{\gamma-1}{\rm exp}\{-b(\lambda y)^c\}[1-{\rm
exp}\{-(\lambda y)^c\}]^{a-1} dy = \frac{\lambda^{-\gamma}}{c}
S_{\frac{\gamma}{c},b,a},
\end{equation}
from which it follows that
$$S_{\gamma/c,b,a} = B(a,b)\lambda^{\gamma-c} E(X^{\gamma-c}),$$
or, equivalently, for any real $r$
\begin{equation}\label{rel2}
E(X^r)=  \frac{1}{\lambda^r B(a,b)} S_{r/c+1,b,a},
\end{equation}
relating $S_{r/c+1,b,a}$ to the $r$th generalized moment of the beta Weibull.

First, we consider the integral (\ref{sdba}) when $d$ is an integer.
Let $U$ be a random variable following the Beta$(b,a)$ distribution
with pdf $f_U(\cdot)$ and $W = -{\rm log}(U)$. Further, let
$F_U(\cdot)$ and $F_W(\cdot)$ be the cdf's of $U$ and $W$,
respectively. It is easy to see that $F_W(x) = 1- F_U(e^{-x})$.
Further, by the properties of the Lebesgue-Stiltjes integral, we
have
\begin{eqnarray*}
E(W^{d-1}) &=& \int_{-\infty}^\infty x^{d-1} dF_W(x)= \int_0^\infty x^{d-1}e^{-x} f_U(e^{-x}) dx\\
           &=& \frac{1}{B(a,b)}\int_0^\infty x^{d-1}e^{-bx}(1-e^{-x})^{a-1}dx\\
           &=& \frac{S_{d,b,a}}{B(a,b)}.
\end{eqnarray*}
Thus, the values of $S_{d,b,a}$ for integer values of $d$ can be
found from the moments of $W$ if they are known. However, the moment
generating function (mgf) $M_W(t) = E(e^{tW})$ of $W$ can be
expressed as
\begin{eqnarray*}
M_W(t) &=& E(U^{-t})= \frac{1}{B(a,b)}\int_0^1 x^{b-t-1} (1-x)^{a-1} dx\\
       &=& \frac{B(b-t,a)}{B(a,b)}.
\end{eqnarray*}
This formula is well defined for $t<b$. However, we are only
interested in the limit $t\rightarrow 0$ and therefore this
expression can be used for the current purpose. We can write
\begin{equation}\label{sinteiro}
S_{d,b,a}=B(a,b) E(W^{d-1}) = B(a,b) M_W^{(d-1)}(0) = \left. \frac{\partial^{d-1}}{\partial t^{d-1}} B(b-t,a)\right\vert_{t=0}.
\end{equation}
From equations (\ref{rel2}) and (\ref{sinteiro}) for any positive integer $k$ we obtain a general formula
\begin{equation}\label{kcmoment}
E(X^{kc}) = \frac{1}{\lambda^{kc} B(a,b)} \left. \frac{\partial^{k}}{\partial t^{k}} B(b-t,a)\right\vert_{t=0}.
\end{equation}
As particular cases, we can see directly from (\ref{sinteiro}) that
$$
S_{1,b,a} = B(a,b), S_{2,b,a} = {B(a,b)\left\{ \psi \left( a+b
\right) -\psi \left( b \right)  \right\}}
$$
and
$$
S_{3,b,a} =  B\left( a,b \right)[ \psi' \left( b \right) -\psi'
\left( a+b \right)   + \left\{  \psi\left( a+b \right)-\psi \left( b
\right)  \right\} ^{2}],
$$
and by using (\ref{rel2}) we find
$$
E(X^c) = \frac{\left\{ \psi \left( a+b \right) -\psi \left( b
\right)  \right\}}{\lambda^c}
$$
and
$$
E(X^{2c}) = \frac{\psi' \left( b \right) -\psi' \left( a+b \right) +
\left\{  \psi\left( a+b \right)-\psi \left( b \right)  \right\}
^{2}}{\lambda^{2c}}.
$$
The same results can also be obtained directly from (\ref{kcmoment}).

Note that the formula for $S_{1,b,a}$ matches the one just given
after equation (12). Since the BW distribution for $c=1$ reduces to
the beta exponential distribution, the above formulae for $E(X^{c})$
and $E(X^{2 c})$ reduce to the corres\-pon\-ding ones obtained by
Nadarajah and Kotz (2006).

Our main goal here is to give the $r$th moment of $X$ for every positive integer $r$. In
fact, in what follows we obtain the $r$th generalized moment for every real $r$ which
may be used for further theoretical or numerical analysis. To this end, we need to obtain a
formula for $S_{d,b,a}$ that holds for every positive real $d$.
In the appendix we show that for any $d>0$ the following identity holds for positive real
non-integer $a$
\begin{equation}\label{identity}
S_{d,b,a} = \Gamma(a)\Gamma\left(d\right)\sum_{j=0}^\infty \frac{(-1)^j}{\Gamma(a-j)j!(b+j)^{d}}
\end{equation}
and that when $a$ is positive integer
\begin{equation}\label{identity2}
S_{d,b,a} = \Gamma\left(d\right)\sum_{j=0}^{a-1}\binom{a-1}{j}\frac{(-1)^j}{(b+j)^{d}}
\end{equation}
is satisfied.

It now follows from (\ref{identity}) and (\ref{rel2}) that the $r$th
generalized moment of $X$ for positive real non-integer $a$ can be
written as
\begin{equation}\label{bwmom}
E(X^r) = \frac{\Gamma(a)\Gamma(r/c +1)}{\lambda^r B(a,b)} \sum_{j=0}^\infty \frac{(-1)^j}{\Gamma(a-j)j!(b+j)^{r/c+1}}.
\end{equation}
When $a>0$ is integer, we obtain
\begin{equation}\label{bwmom2}
E(X^r) = \frac{\Gamma(r/c +1)}{\lambda^r B(a,b)} \sum_{j=0}^{a-1}\binom{a-1}{j} \frac{(-1)^j}{(b+j)^{r/c+1}}
\end{equation}
When $a=b=1$, $X$ follows a Weibull distribution and (\ref{bwmom2})
becomes
$$E(X^r) = \frac{\Gamma(r/c + 1)}{\lambda^r},$$
which is precisely the $r$th moment of a Weibull distribution with
parameters $\lambda$ and $c$. Equations (\ref{kcmoment}),
(\ref{bwmom}) and (\ref{bwmom2}) represent the main results of this
section, which may serve as a starting point for applications for
particular cases as well as further research.\\

\begin{figure}[!h]\includegraphics[width=5.0in]{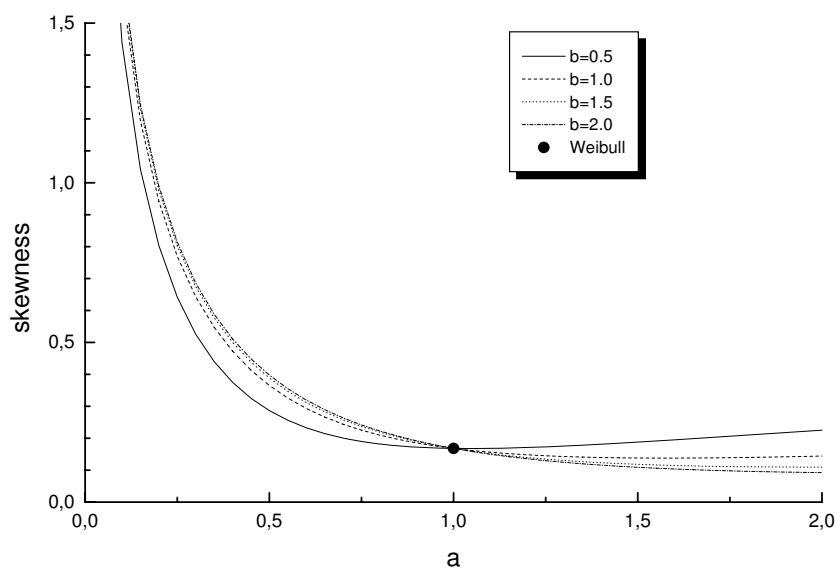}
\caption{\label{fig3} Skewness of the BW distribution as a function
of parameter $a$, for several values of parameter $b$ }
\end{figure}

\begin{figure}[!h]\includegraphics[width=5.0in]{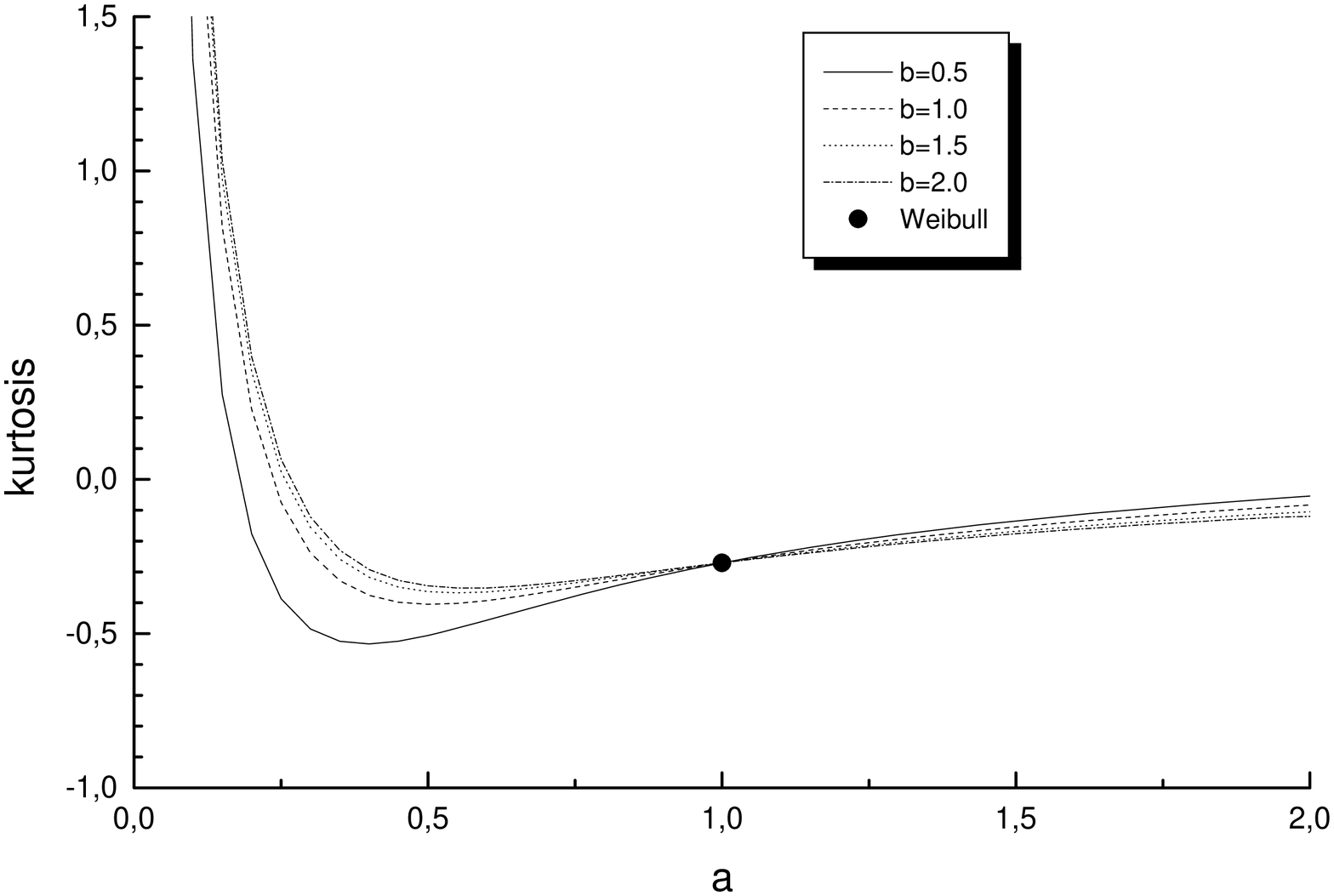}
\caption{\label{fig4} Kurtosis of the BW distribution as a function
of parameter $a$, for several values of parameter $b$ }
\end{figure}

\begin{figure}[!h]\includegraphics[width=5.0in]{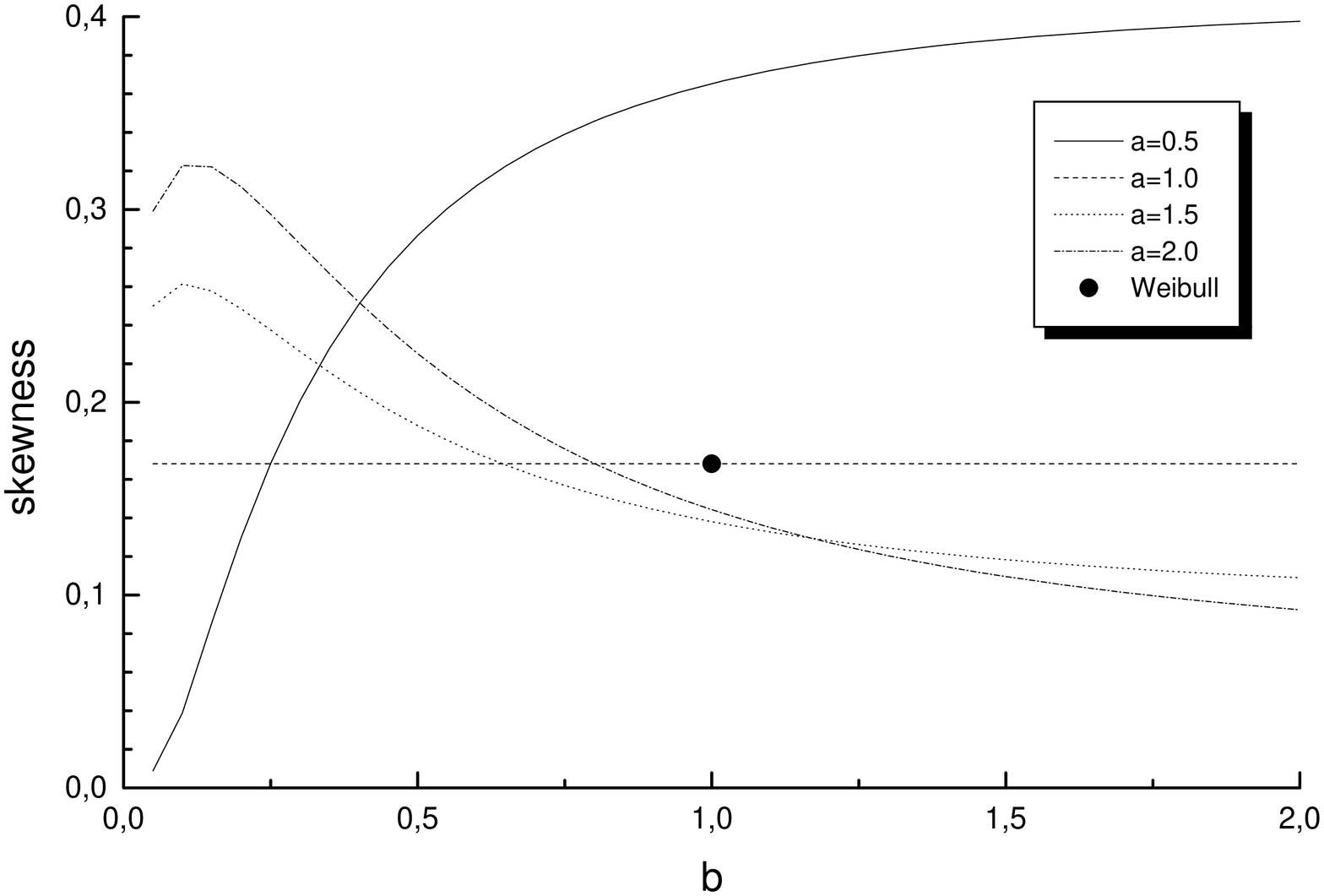}
\caption{\label{fig5} Skewness of the BW distribution as a function
of parameter $b$, for several values of parameter $a$ }
\end{figure}

\begin{figure}[!h]\includegraphics[width=5.0in]{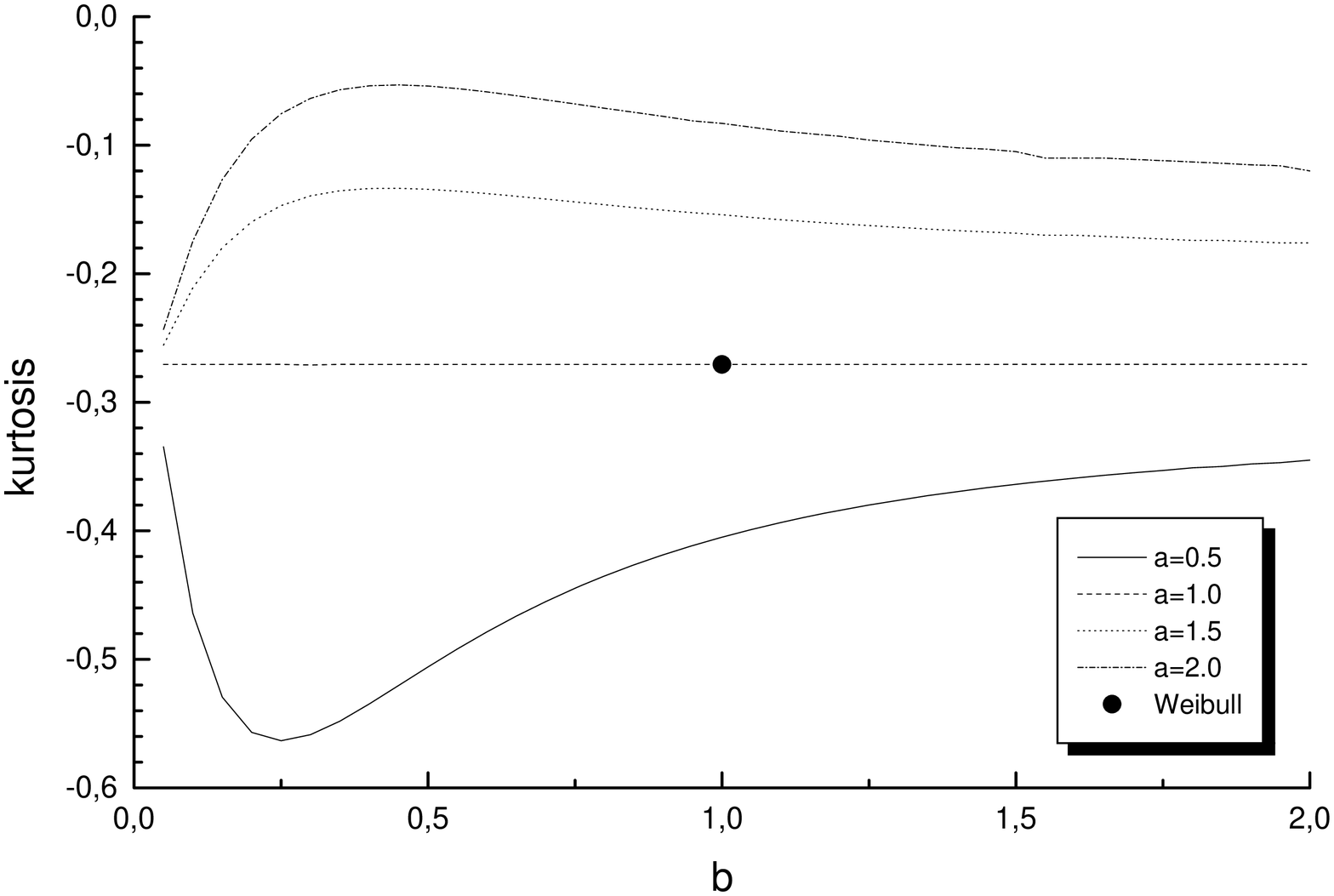}
\caption{\label{fig6} Kurtosis of the BW distribution as a function
of parameter $b$, for several values of parameter $a$ }
\end{figure}

Graphical representation of skewness and kurtosis for some choices
of parameter $b$ as function of parameter $a$, and for some choices
of parameter $a$ as function of parameter $b$, for fixed $\lambda=1$
and $c=3$, are given in Figures \ref{fig3} and \ref{fig4}, and
\ref{fig5} and \ref{fig6}, respectively. It can be observed from
Figures 3 and 4 that the skewness and kurtosis curves cross at
$a=1$, and from Figures 5 and 6 that both skewness and kurtosis are
independent of $b$ for $a=1$. In addition, it should be noted that
the Weibull distribution (equivalent to BW for $a=b=1$) is
represented by as single point on Figures 3-6.

\section{Moment Generating Function}

We can give an expansion for the mgf of the BW distribution as
follows
\begin{eqnarray*}
 M(t)&=& \frac{c\lambda^c}{B(a,b)}\int_0^\infty {\rm exp}(tx)x^{c-1} {\rm exp}\{-b(\lambda x)^c\}[1-{\rm exp}\{-(\lambda x)^c\}]^{a-1} dx\\
 &=&\frac{c\lambda^c}{B(a,b)}\sum_{r=0}^\infty \frac{t^r}{r!} \int_0^\infty x^{r+c-1} {\rm exp}\{-b(\lambda x)^c\}[1-{\rm exp}\{-(\lambda x)^c\}]^{a-1} dx\\
 &=&\frac{c\lambda^c}{B(a,b)}\sum_{r=0}^\infty \frac{t^r}{r!} \frac{\lambda^{-(r+c)}}{c}S_{r/c+1,b,a},
 \end{eqnarray*}
 where the last expression comes from (\ref{rel1}). For positive real
 non-integer $a$ using (\ref{identity}) we have
 \begin{equation}\label{bwmgf}
 M(t) = \frac{\Gamma(a)}{B(a,b)}\sum_{r=0}^\infty \sum_{j=0}^\infty \frac{t^r \Gamma(r/c +1) (-1)^j}{\lambda^r \Gamma(a-j)(b+j)^{r/c +1}r!j!},
 \end{equation}
 and for integer $a>0$ using (\ref{identity2}) we obtain
 \begin{equation}\label{bwmgf2}
 M(t) = \frac{1}{B(a,b)}\sum_{r=0}^\infty \frac{t^r\Gamma(r/c +1)}{\lambda^r r!} \sum_{j=0}^{a-1}\binom{a-1}{j} \frac{(-1)^j}{(b+j)^{r/c +1}}.
 \end{equation}
 Note that the expression for the mgf obtained by Choudhury (2005) is a particular case of (\ref{bwmgf}), when $a=\theta$, $\lambda=1/\alpha$ and $b=1$.
 When $c=1$, we have
 \begin{eqnarray*}
 M(t)&=&\sum_{r=0}^\infty \frac{\lambda^{-r}t^r}{cB(a,b) r!} S_{r+1,b,a}\\
 &=&\frac{\lambda}{B(a,b)}\int_0^\infty e^{tx-b\lambda x}(1-e^{-\lambda x})^{a-1} dx.
 \end{eqnarray*}
Substituting $y={\rm exp}(-\lambda x)$ in the above integral yields
\begin{equation}\label{bemgf}
M(t) = \frac{1}{B(a,b)}\int_0^1 y^{b-t/\lambda -1} (1-y)^{a-1}dy
\end{equation}
and using the definition of the beta function in (\ref{bemgf}) we
find
$$M(t) = \frac{B(b-t/\lambda,a)}{B(a,b)},$$
which is precisely the expression (3.1) obtained by Nadarajah and
Kotz (2006).

\section{Estimation and information matrix}
Let $Y$ be a random variable with the BW distribution (\ref{bwpdf}).
The log-likelihood for a single observation $y$ of $Y$ is given by
\begin{eqnarray*}
\ell(\lambda,c,a,b) &=& {\rm log}(c) + c\,{\rm log}(\lambda) + (c-1){\rm log}(y) - {\rm log}\left\{B(a,b)\right\} - b (\lambda y)^c\\
&+& (a-1){\rm log}[1-{\rm exp}\{-(\lambda y)^c\}].
\end{eqnarray*}
The corresponding components of the score vector are:
\begin{equation}\label{score3}
\frac{\partial \ell}{\partial a} = - \left\{ \psi \left( a \right)
-\psi \left( a+b \right)  \right\} + \log  \left\{ 1-{e^{- \left(
\lambda\,y \right) ^{c}}} \right\},
\end{equation}
\begin{equation}\label{score4}
\frac{\partial \ell}{\partial b} = - \left\{ \psi \left( b \right)
-\psi \left( a+b \right)  \right\} - \left( \lambda\,y \right) ^{c},
\end{equation}
\begin{eqnarray}
\frac{\partial \ell}{\partial c} = {\frac {1}{c}}
+\log  \left( \lambda\,y \right)
-b \left( \lambda\,y \right) ^{c}\log  \left( \lambda\,y \right)
+ {\frac { \left( a-1 \right)  \left( \lambda\,y
 \right) ^{c}\log  \left( \lambda\,y \right) {e^{- \left( \lambda
\,y \right) ^{c}}}}{1-{e^{- \left( \lambda\,y \right) ^{c}
}}}}\label{score2}
\end{eqnarray}
and
\begin{equation}\label{score1}
\frac{\partial \ell}{\partial \lambda} = {\frac {c}{\lambda}}-
{\frac {b\,c }{\lambda}}\left( \lambda\,y \right) ^{c}
+\frac { c\left( a-1 \right)  \left( \lambda\,y
 \right) ^{c}{e^{- \left( \lambda\,y \right) ^{c}}}}
 {\lambda\, \left\{ 1-{e^{- \left( \lambda\,y \right) ^{c}}} \right\} }.
\end{equation}
The maximum likelihood equations derived by equating
(\ref{score3})-(\ref{score1}) to zero can be solved numerically for
$a, b, c$ and $\lambda$. We can use iterative techniques such as a
Newton-Raphson type algorithm to obtain the estimates of these
parameters. It may be worth noting from $E(\partial \ell/\partial b)
= 0$, that (\ref{score4}) yields
$$E(Y^c) = \frac{ \psi \left( a+b \right) -\psi \left( b \right)
}{\lambda^c},$$
which agrees with the previous calculations.

For interval estimation of $(a,b,c,\lambda)$ and hypothesis tests,
the Fisher information matrix is required. For expressing the
elements of this matrix it is convenient to introduce an extension
of the integral (\ref{sdba})
\begin{eqnarray}\label{tdbae}
T_{d,b,a,e}=
\int_0^{\infty}{ x^{d-1} e^{-b x} \left(1-e^{-x} \right)^{a-1}
\left(\log x\right)^e dx},
\end{eqnarray}
so that we have
\begin{eqnarray}
T_{d,b,a,0}=
S_{d,b,a}.\nonumber
\end{eqnarray}
As before, let $W = -\log(U)$, where $U$ is a random variable following the Beta$(b,a)$ distribution, then
\begin{eqnarray*}
E[W^{d-1}\{\log(W)\}^e] &=& \int_{-\infty}^\infty x^{d-1} (\log x)^e dF_W(x)\\
                        &=& \frac{1}{B(a,b)}\int_0^\infty x^{d-1} e^{bx} (1-e^{-x})^{a-1}(\log x)^e dx\\
                        &=& \frac{T_{d,b,a,e}}{B(a,b)}.
\end{eqnarray*}
Hence, the equation
\begin{equation}\label{relt}
T_{d,b,a,e} = B(a,b) E[W^{d-1}\{\log(W)\}^e],
\end{equation}
relates $T_{d,b,a,e}$ to expected values.

To simplify the expressions for some elements of the information matrix, it is useful to note the identities
$$S_{1,b+2,a} -2 S_{1,b+1,a} + S_{1,b,a}=B(a+2,b)$$
and
$$b S_{2,b,a} - (a+2b+1) S_{2,b+1,a} + (a+b+1) S_{2,b+2,a}=B(a+2,b),$$

\noindent
which can be easily proved.

Explicit expressions for the elements of the information matrix $K$,
obtained using Maple and Mathematica algebraic manipulation software
(we have used both for double checking the obtained expressions),
are given below in terms of the integrals (\ref{sdba}) and
(\ref{tdbae}):

\begin{eqnarray*}
\kappa_{a,a}= \psi' \left( a \right) -\psi' \left( a+b \right),
\end{eqnarray*}
\begin{eqnarray*}
\kappa_{a,b}= -\psi' \left( a+b \right),~ \kappa_{a,c}= -{\frac
{T_{{2,b+1,a-1,1}}}{c B \left( a,b \right) }},~ \kappa_{a,\lambda}=
-{\frac {c\,S_{{2,b+1,a-1}}}{\lambda B  \left( a,b \right) }} ,
\end{eqnarray*}
\begin{eqnarray*}
\kappa_{b,b}= \psi' \left( b \right) -\psi' \left( a+b \right),~
\kappa_{b,c}= {\frac {T_{{2,b,a,1}}}{cB  \left( a,b \right) }},~
\kappa_{b,\lambda}= \frac {c}{\lambda } {\frac {S_{{2,b,a}}}{B
\left( a,b \right) }} ,
\end{eqnarray*}

\begin{eqnarray*}
\kappa_{c,c}=&&
\frac {1}{{c}^{2}}
+
\frac{1}{c^2\,B  \left( a,b \right)}
\left\{
 \left( a-1 \right) T_{{3,b+1,a-2,2}}
\right.\nonumber\\
 &&\left.
+b\,T_{{2,b,a-2,2}}-\left( a+2\,b
-1\right) T_{{2,b+1,a-2,2}}
+ \left( a+b-1 \right) T_{{2,b+2,a-2,2}}
\right\},
\end{eqnarray*}

\begin{eqnarray*}
\kappa_{c,\lambda}=&&
\frac {1}{\lambda\,B  \left( a,b \right) }
\left\{
\left( a-1 \right) T_{{3,b+1,a-2,1}}
\right.
\nonumber\\
&&\left.
+b\,T_{{2,b,a-2,1}}
- \left(a+2\,b -1\right) T_{{2,b+1,a-2,1}}
+\left( a+b-1 \right) T_{{2,b+2,a-2,1}}
\right\}
\end{eqnarray*}

and

\begin{eqnarray*}
\kappa_{\lambda,\lambda}=&& \frac {c^2}{{\lambda}^{2}} +\frac
{{c}^{2} \left( a-1 \right)}{{\lambda}^{2}B  \left( a,b \right) }
S_{{3,b+1,a-2}}.
\end{eqnarray*}
The integrals $S_{{i,j,k}}$ and $T_{{i,j,k,l}}$ in the information
matrix are easily numerically determined using MAPLE and MATHEMATICA
for any $a$ and $b$.

Under conditions that are fulfilled for parameters in the interior
of the parameter space but not on the boundary, the asymptotic
distribution of the maximum likelihood estimates $\hat{a}, \hat{b},
\hat{c}$ and $\hat{\lambda}$ is multivariate normal $N_4(0,K^{-1})$.
The estimated multivariate normal $N_4(0,\widehat{K}^{-1})$
distribution can be used to cons\-truct approximate confidence
intervals and confidence regions for the individual parameters and
for the hazard rate and survival functions. The asymptotic normality
is also useful for testing goodness of fit of the BW distribution
and for comparing this distribution with some of its special
sub-models using one of the three well-known asymptotically
equivalent test statistics - namely, the likelihood ratio (LR)
statistic, Wald and Rao score statistics.

We can compute the maximum values of the unrestricted and restricted
log-likelihoods to construct the LR statistics for testing some
sub-models of the BW distribution. For example, we may use the LR
statistic to check if the fit using the BW distribution is
statistically ``superior'' to a fit using the exponentiated Weibull
or Weibull distributions for a given data set. Mudholkar et al.
(1995) in their discussion of the classical bus-motor-failure data,
noted the curious aspect in which the larger EW distribution
provides an inferior fit as compared to the smaller Weibull
distribution.

\section{Application to real data}

In this section we compare the results of fitting the BW and Weibull distribution
to the data set studied by Meeker and Escobar (1998, p. 383), which gives the times of failure and running times for a sample of devices from a field-tracking study of
a larger system. At a certain point in time, 30 units were installed in normal
service conditions. Two causes of failure were observed for each unit
that failed: the failure caused by an accumulation of randomly occurring
damage from power-line voltage spikes during electric storms and failure
caused by normal product wear. The times are: 275, 13, 147, 23, 181, 30, 65, 10, 300, 173, 106, 300, 300, 212, 300, 300, 300, 2, 261, 293, 88, 247, 28, 143, 300, 23, 300, 80, 245, 266.

The maximum likelihood estimates and the maximized log-likelihood $\hat{l}_{BW}$ for the BW distribution are:
$$\hat{a} = 0.0785, \hat{b} = 0.0659, \hat{c} = 7.9355, \hat{\lambda} = 0.004987
\hbox{~~and~~}\hat{l}_{BW} = -169.919,$$
while the maximum likelihood estimates and the maximized log-likelihood $\tilde{l}_W$ for the Weibull distribution are:
$$ \tilde{c} = 1.2650, \tilde{\lambda} = 0.005318 \hbox{~~and~~} \tilde{l}_W = -184.3138.$$

The likelihood ratio statistic for testing the hypothesis $a=b=1$ (namely, Weibull versus
BW distribution) is then $w=28.7896$, which indicates that the Weibull distribution should be rejected. As an alternative test we use the Wald statistic. The asymptotic
covariance matrix of the maximum likelihood estimates for the BW distribution,
which comes from the inverse of the information matrix, is given by
$$
\widehat{K}^{-1}=
10^{-7}\times
\left(
\begin{array}{rrrr}
 8699.35364 & 4743.69977 & -488130.870
   & 87.9136383 \\
 4743.69977 & 13079.4394 & -4009.69885
   & -135.603333 \\
 -488130.870 & -4009.69885 & 58517447.8 &
   -16222.8149 \\
 87.9136383 & -135.603333 & -16222.8149 &
   6.19530131
\end{array}
\right).
$$
The resulting Wald statistic is found to be $W=38.4498$, again signalizing that
the BW distribution conform to the above data. In Figure 7 we display the pdf of both
Weibull and BW distributions fitted and the data set, where it is seen that the
BW model captures the aparent bimodality of the data.

\begin{figure}[!h]\includegraphics[width=5.0in]{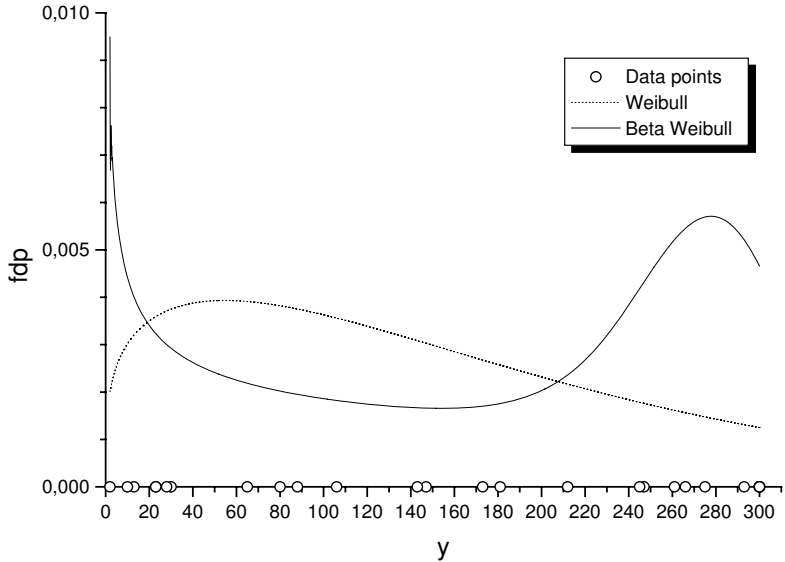}
\caption{\label{fig7}
The probability density function (\ref{bwpdf})
of the fitted BW and Weibull distributions}
\end{figure}

\section{Conclusion}

The Weibull distribution, having exponential and Rayleigh as special
cases, is a very popular distribution for modeling lifetime data and
for modeling phenomenon with monotone failure rates. In fact, the BW
distribution represents a generalization of several distributions
previously considered in the literature such as the exponentiated
Weibull distribution (Mudholkar et al., 1995, Mudholkar and Hutson,
1996, Nassar and Eissa, 2003, Nadarajah and Gupta, 2005 and
Choudhury, 2005) obtained when $b=1$. The Weibull distribution (with
parameters $c$ and $\lambda$) is also another particular case for
$a=1$ and $b=1$. When $a=1$, the BW distribution reduces to a
Weibull distribution with parameters $\lambda \, b^{1/c}$ and $c$.
The beta exponential distribution is also an important special case
for $c=1$.

The BW distribution provides a rather general and flexible framework
for statistical analysis. It unifies several previously proposed
families of distributions, therefore yielding a general overview of
these families for theoretical studies, and it also provides a
rather flexible mechanism for fitting a wide spectrum of real world
data sets.

We derive explicit expressions for the moments of the BW
distribution, including an expansion for the moment generating
function. These expressions are manageable and with the use of
modern computer resources with analytic and numerical capabilities,
may turn into adequate tools comprising the arsenal of applied
statisticians. We discuss the estimation procedure by ma\-xi\-mum
likelihood and derive the information matrix. Finally, we
demonstrate an application to real data.

\section*{Appendix}
In what follows, we derive the identities (\ref{identity}) and (\ref{identity2}).
We start from
$$f(x) = {\rm exp}(-bx)(1-e^{-x})^{a-1},$$
which yields
$$\int_0^\infty x^{d -1} f(x)dx = \int_0^\infty x^{d-1} {\rm exp}(-bx)(1-e^{-x})^{a-1} dx,$$
and substituting $z = e^{-x}$ gives
\begin{equation}\label{rec}
\int_0^\infty x^{d-1} f(x)dx = \int_0^1 \vert{\rm log}z\vert^{d-1} z^{b-1}(1-z)^{a-1} dz.
\end{equation}
For real non-integer $a$, we have
$$ \int_0^\infty x^{d-1} f(x)dx = \Gamma(a)\sum_{j=0}^\infty \frac{(-1)^j}{\Gamma(a-j)j!}\int_0^1\vert{\rm log}z\vert^{\gamma/c-1}z^{b+j-1} dz.$$
Also, for real $p>-1$ and real $q$, we have
\begin{equation}\label{integral}
\int_0^1 x^p \vert {\rm log}x\vert^q dx = \frac{\Gamma(1+q)}{(1+p)^{q+1}}.
\end{equation}
Hence,
$$ \int_0^\infty x^{d-1} f(x)dx = \Gamma(a)\sum_{j=0}^\infty \frac{(-1)^j\Gamma(d)}{\Gamma(a-j)j!(b+j)^{d}},$$
and, finally, we arrive at
$$\int_0^\infty x^{d -1}{\rm exp}(-bx)(1-e^{-x})^{a-1} dx = \Gamma(a)\Gamma\left(d\right)\sum_{j=0}^\infty \frac{(-1)^j}{\Gamma(a-j)j!(b+j)^{d}},$$
which represents the identity (\ref{identity}).

Now, let $a>0$ be an integer; then, from (\ref{rec}), we have
$$ \int_0^\infty x^{d-1} f(x)dx = \sum_{j=0}^{a-1} \binom{a-1}{j}{(-1)^j}\int_0^1\vert{\rm log}z\vert^{d-1}z^{b+j-1} dz.$$
Using (\ref{integral}) we obtain
$$\int_0^\infty x^{d-1} f(x)dx = \sum_{j=0}^{a-1}\binom{a-1}{j}\frac{(-1)^j\Gamma(d)}{(b+j)^{d}},$$
and therefore we arrive at
$$\int_0^\infty x^{d -1}{\rm exp}(-bx)(1-e^{-x})^{a-1} dx = \Gamma\left(d\right)\sum_{j=0}^{a-1}\binom{a-1}{j} \frac{(-1)^j}{(b+j)^{d}},$$
which represents the identity (\ref{identity2}).

\end{document}